\def\e{{\rm e}}
\def\del{\partial}
\def\half{{1\over2}}

\def\vev#1{\langle #1 \rangle}
\def\del{\partial}
\def\half{{1\over2}}

\def\vev#1{\langle #1 \rangle}

\def\del{\partial}
\def\dslash{\del\kern-0.55em\raise 0.14ex\hbox{/}}

\def\tLambda{\tilde\Lambda}
\def\tpe{\tilde p_E}
\def\rough#1{\raise.3ex\hbox{$#1$\kern-.75em\lower1ex\hbox{$\sim$}}}

\newcommand{\PRD}[3]{{\it Phys. Rev.} {\bf D{#1}}, {#2} (19{#3})}

\newcommand{\PRDM}[3]{{\it Phys. Rev.} {\bf D{#1}}, {#2} (20{#3})}

\newcommand{\NPB}[3]{{\it Nucl. Phys.} {\bf B{#1}}, {#2} (19{#3})}
\newcommand{\NPBM}[3]{{\it Nucl. Phys.} {\bf B{#1}}, {#2} (20{#3})}
\newcommand{\PLB}[3]{{\it Phys. Lett.} {\bf B{#1}}, {#2} (19{#3})}
\newcommand{\PLBM}[3]{{\it Phys. Lett.} {\bf B{#1}}, {#2} (20{#3})}

\newcommand{\PTPM}[3]{{\it Prog. Theor. Phys.} {\bf {#1}}, {#2} (20{#3})}
\newcommand{\ANN}[3]{{\it Ann. Phys. (N.Y.)} {\bf {#1}}, {#2} (19{#3})}

\newcommand{\MPL}[3]{{\it Mod. Phys. Lett.} {\bf A{#1}} {#2} (19{#3})}
\newcommand{\MPLM}[3]{{\it Mod. Phys. Lett.} {\bf A{#1}} {#2} (20{#3})}

\newcommand{\jhep}[3]{{\it JHEP} {\bf {#1}} (20{#2}) {#3}}

\newcommand{\hmu}{\hat\mu}

\documentclass[12pt]{article}
\usepackage{amsmath}
\usepackage{amssymb}
\usepackage{graphicx}
\begin{document}
\baselineskip=18pt
\begin{titlepage}
\begin{flushright}
\end{flushright}
\vspace{1cm}
\begin{center}{\Large\bf 
Is the  Higgs a sign of extra dimensions ?
}
\end{center}
\vspace{1cm}
\begin{center}
Hiroto So$^{(a)}$
\footnote{E-mail: so@phys.sci.ehime-u.ac.jp} and
Kazunori Takenaga$^{(b)}$
\footnote{E-mail: takenaga@kumamoto-hsu.ac.jp}
\end{center}
\vspace{0.2cm}
\begin{center}
${}^{(a)}$ {\it Department of Physics, Ehime University, 
Bunkyou-chou 2-5, Matsuyama 790-8577, Japan}
\\[0.2cm]
${}^{(b)}$ {\it Faculty of Health Science, Kumamoto
Health Science University, Izumi-machi, Kita-ku, Kumamoto 861-5598, Japan}
\end{center}
\vspace{1cm}
\begin{abstract}
We introduce a $4$-dimensional cutoff in the scenario of gauge-Higgs
unification to control the ultraviolet behavior. A one-loop 
effective potential for a Higgs field and the Higgs mass are
obtained with the cutoff. We find an 
{\it interrelation} between the $4$-dimensional cutoff and the scale of extra 
dimensions, which is concretized through the Higgs mass. Combining this
interrelation and the recently discovered Higgs boson at LHC, we obtain 
an interesting constraint for the $4$-dimensional 
cutoff and the extra dimensional scale.
%
%
\end{abstract}
\end{titlepage}
\newpage
\section{Introduction}
A higher dimensional gauge theory is one of the attractive candidates 
for physics beyond the standard model. Gauge-Higgs unification \cite{manton} is one
of such the gauge theories, where gauge and Higgs fields are unified into
a higher dimensional gauge field. Component gauge fields
for compactified extra directions behave like the Higgs fields at low energy.

In the scenario of the gauge-Higgs unification, the gauge symmetry is broken
through quantum corrections \cite{hosotani}, and the Higgs mass, which is zero at the
tree-level due to the higher dimensional gauge invariance, arises 
at quantum level. It has been said that the effective potential for the Higgs field 
and the Higgs mass do not suffer from ultraviolet divergences. Thanks to this
property, the gauge-Higgs unification may solve the gauge hierarchy
problem without relying on supersymmetry \cite{lim}. The gauge-Higgs 
unification has been an attractive alternative for the Higgs mechanism.
Many attempts to seek phenomenologically viable models with 
the gauge-Higgs unification have been done in the 
past \cite{gaugehiggs, models, model2}. In addition to it, various aspects 
of the gauge-Higgs unification such as 
the finite temperature phase transition {\it et al.} have also 
been studied \cite{finitet, renormalization, lat, hst}.

In the gauge-Higgs unification, one needs to evaluate the effective 
potential for the Higgs field in order to discuss
the gauge symmetry breaking patterns and to calculate the Higgs mass which is obtained
by the second derivative of the potential at the vacuum. In the past, one employed 
the dimensional regularization for the momentum integration in
evaluating the effective potential at the one-loop level. The divergent terms 
that depend on the order parameter (the Higgs field) do not appear in the effective 
potential and the Higgs mass. But the dimensional regularization essentially 
can not account for power divergences.

As stated above, the Higgs mass arises through quantum corrections in extra
dimensions, say, Kaluza-Klein modes in the gauge-Higgs 
unification. It is, however, difficult to obtain the definite quantum effect of  
the higher dimensional gauge theory because of the nonrenormalizability.
The detailed structure of the effective potential for the Higgs field is unknown
as long as one can not solve the dynamics in higher dimensions. 
At the moment, it remains unclear how much one should take the quantum correction 
in the extra dimension into account in order to determine the low-energy physics.

The effective potential we shall compute has
the Kaluza-Klein modes and the $4$-dimensional momentum 
cutoff which is originated from the $5$-dimensional cutoff because we start with 
the $5$-dimensional gauge theory in which there are uncontrollable ultraviolet divergences
due to the nonrenormalizability. We would like to keep the shift symmetry \cite{shift} which is 
a remnant of the original gauge symmetry, so that one 
has to sum up all the Kaluza-Klein modes\footnote{When we consider an effective theory of the
$5$-dimensional gauge theory with the cutoff, it is natural to respect the shift 
symmetry as the $4$-dimensional theory.}. 
Then the $5$-dimensional ultraviolet divergence 
reduces to the $4$-dimensional momentum cutoff.

In the theory like the gauge-Higgs unification, the $5$-dimensional
physics, the Kaluza-Klein mode  determines the low-energy physics, for example, the 
Higgs mass. It is important to have the parameter which tells us how much 
the $5$-dimensional physics contributes to determine the low-energy physics. 
Such the parameter can be constructed by using the $4$-dimensional momentum 
cutoff and the $5$-dimensional scale in our case. We shall call the parameter as
the {\it interrelation}. It should be noted that the 
interrelation is not a phenomenological parameter, but is a theoretical one. 
It is interesting, however, that if one takes account of the experimental value of
the physical observable such as the Higgs mass, one obtains a constraint on the
interrelation by which we understand how much the Kaluza-Klein mode should contribute
to the Higgs mass.

This paper is organized as follows. In the next section, after brief setup, we present the 
expression for one-loop effective potential for the Higgs field and the Higgs mass 
with the $4$-dimensional cutoff. The interrelation, which is a key notion, is also explained.
We also find that there is a remarkable combination between the periodicity of the Higgs 
field and an exponential suppression with respect to the interrelation. In 
section $3$, we study the interrelation through the Higgs mass in some models 
with the gauge-Higgs unification. We give a constraint on the $4$-dimensional cutoff
and the scale the extra dimension by taking account of the result on the Higgs 
mass at LHC \cite{lhc1, lhc2}. The final section 
is devoted to the conclusions. In Appendix, important formulae used on the text are derived.
\section{Effective potential and Higgs mass with $4$-dimensional cutoff}
Let us consider a nonsupersymmetric $SU(3)$ gauge theory 
on $M^4\times S^1/Z_2$, where $M^4$ is the $4$-dimensional 
Minkowski space-time and $S^1/Z_2$ is an 
orbifold\footnote{Notations used in this paper are the same as
those in \cite{marutake}.}. One must specify boundary conditions
of fields for the $S^1$ direction and the two orbifold 
fixed points at $y=0, \pi R$, where $R$ is 
the radius of the $S^1$. They are defined by
\begin{eqnarray}
A_{\hmu}(x^{\mu}, y+2\pi R)&=&UA_{\hmu}(x^{\mu}, y)U^{\dagger},\\
\begin{pmatrix}
A_{\mu}\\
A_y
\end{pmatrix}(x^{\mu}, y_i-y)
&=&P_i 
\begin{pmatrix}
A_{\mu}\\
-A_y
\end{pmatrix}(x^{\mu}, y_i+y)P_i^{\dagger}\qquad (i=0, 1),
\label{shiki1}
\end{eqnarray}
where $U=U^{\dagger}, P_i^{\dagger}=P_i=P_i^{-1}$ and $y_0=0, y_1=\pi R$. 
The coordinate $x^{\mu} (\mu=0, \cdots, 3)$ denotes the $4$-dimensional
Minkowski space time and $y$ is the coordinate of the extra dimension. 
Since the translation $U$ together with the reflection $P_1$ is equivalent to the 
reflection $P_0$, so that there is a relation $U=P_1P_0$. We 
take $P_i (i=0,1 )$ to be fundamental projections.

In the scenario of the gauge-Higgs unification, the zero modes 
for $A_y$ play an important role and behave Higgs fields 
at low energy. If the Higgs field develops the vacuum expectation value, the
$SU(2)\times U(1)$ gauge symmetry is broken to the electromagnetic $U(1)_{em}$. One 
must choose the boundary 
conditions $P_{0, 1}$ in such a way that the zero mode for $A_y$ belongs to the 
fundamental representation under the $SU(2)$ gauge group. We choose
$P_0=P_1={\rm diag.}(-1, -1, 1)$. Then the $SU(3)$ gauge symmetry is 
broken explicitly down to $SU(2)\times U(1)$ by the orbifolding. 
The zero modes for the gauge field are read off by Eq. (\ref{shiki1}) for the 
boundary condition by $P_{0, 1}$.

The zero modes for $A_{\mu}$ are given by
\begin{equation}
A_{\mu}^{(0)}=\half 
\begin{pmatrix}
A_{\mu}^3+{A_{\mu}^8\over\sqrt{3}} & A_{\mu}^1-iA_{\mu}^2& 0 \\
A_{\mu}^1+iA_{\mu}^2 & -A_{\mu}^3+ {A_{\mu}^8\over\sqrt{3}} & 0\\
0 &0 & -{2\over\sqrt{3}}A_{\mu}^8
\end{pmatrix},
\end{equation}
by which the residual gauge symmetry is clearly $SU(2)\times U(1)$. 
On the other hand, the zero mode for $A_y$ is found to be
\begin{equation}
A_y^{(0)}=\half 
\begin{pmatrix}
0 & 0& A_y^4-iA_y^5 \\[0.1cm]
0 & 0 & A_y^6-iA_y^7\\[0.1cm]
A_y^4+iA_y^5 & A_y^6+iA_y^7 & 0
\end{pmatrix}.
\end{equation}
We observe that 
\begin{equation}
\Phi\equiv \sqrt{2\pi R}~{1\over\sqrt{2}}
\begin{pmatrix}
A_y^4-iA_y^5\\[0.1cm]
A_y^6-iA_y^7
\end{pmatrix}
\end{equation}
belongs to the fundamental representation under the $SU(2)$. 
The adjoint representation of the $SU(3)$ is decomposed under the 
$SU(2)$ into 
\begin{equation}
{\bf 8}\rightarrow {\bf 3}+{\bf 2}+{\bf 2}^{*}+{\bf 1}.
\end{equation}
We understand how the gauge and Higgs fields are embedded into
the higher dimensional gauge field.

By utilizing the $SU(2)\times U(1)$ degrees of freedom, 
the vacuum expectation value for the Higgs field is parametrized by 
\begin{equation}
\vev{A_y^{6}}={a\over g R},
\end{equation}
where $g$ is the $5$-dimensional gauge coupling and $a$ is a real parameter. 
The parameter $a$ is related with the Wilson line phase,
\begin{equation}
W={\cal P}{\rm exp}\left(ig\oint_{S^1} dy \vev{A_y}\right)
=\begin{pmatrix}
1 & 0 & 0 \\
0 &\cos(\pi a) &i\sin(\pi a) \\
0 & i\sin(\pi a) & \cos(\pi a) \\
\end{pmatrix}\quad (a~\mbox{mod}~2).
\end{equation}
The original gauge invariance, namely concerning with the fifth direction,  guarantees 
that the Wilson line phase is mod $2$. The gauge 
symmetry breaking patterns of the $SU(2)\times U(1)$ 
are classified by the values of $a$, 
\begin{equation}
SU(2)\times U(1)\rightarrow \left\{
\begin{array}{ll}
SU(2)\times U(1) & \mbox{for}~~ a=0,\\ 
U(1)\times U(1)'& \mbox{for}~~a=1,\\
U(1)_{\rm em} & \mbox{for~~otherwise}.
\end{array}\right.
\end{equation}
The value of $a$ is determined as the global minimum of the effective potential
for the Higgs field.

In the scenario of the gauge-Higgs unification, one needs not 
only the matter fields that satisfy the periodic 
boundary condition (PBC), but also the ones that satisfy the antiperiodic
boundary condition (APBC). They are distinguished by the 
parameter $\eta ( =1~\mbox{for PBC}, -1~\mbox {for  APBC})$ \cite{marutake}.
In addition to them, we also consider the matter fields belonging to the 
large representation under the $SU(3)$ gauge group such as the adjoint
representation. These are necessary ingredients for the viable model with
the gauge-Higgs unification.

In a gauge-Higgs unification scenario, we start with the 
following 5-dimensional effective potential contribution,
\begin{equation}
F_5(Qa,\delta,\Lambda)= {1\over 4\pi
R}\sum_{n=-\infty}^{\infty}\int^{\Lambda}_{-\Lambda}
{d^4p\over (2\pi)^4}~{\rm ln} \left[p_E^2+\left({n+Qa-{\delta\over 2}\over R}\right)^2\right],
\label{neweq}
\end{equation}
where $Q=1, 1/2$ for the adjoint, fundamental representation under
the $SU(2)$, respectively. The parameter $\delta$ takes $0~(1)$ for the
field with the PBC (APBC). We have introduced the $4$-dimensional ultraviolet
cutoff $\Lambda$ in the momentum integration originated in $5$-dimensional 
ultraviolet cutoff because our starting theory is a $5$-dimensional 
Yang-Mills theory and has some ultraviolet-divergent 
quantities owing to the nonrenormalizability. Noting that it is necessary 
to sum up all the Kaluza-Klein modes in order to keep the shift 
invariance reflected as $5$-dimensional gauge invariance, the 5-dimensional 
ultraviolet divergence reduces to the $4$-dimensional cutoff $\Lambda$, (\ref{neweq}).
The effective potential is given by collecting all the contributions of 
the fields in theory,
\begin{equation}
V_{eff}=\sum_{i=fields}(-1)^F N_{deg}^i ~F_5^i(Qa, \delta).
\label{shiki9}
\end{equation}
The $F$ stands for the fermion number of the internal loop, and $N_{deg}^i$ is 
the number of on-shell degrees of freedom for the relevant matter field.

We first sum up all the Kaluza-Klein modes, 
\begin{equation}
\sum_{n=-\infty}^{\infty} {2p_E R^2\over (Rp_E)^2+(n+Qa-{\delta \over 2})^2}
=L\times {\sinh(Lp_E)\over \cosh(Lp_E)-\cos(2\pi (Qa-{\delta \over 2}))},
\label{shiki7}
\end{equation}
where we have defined $L\equiv 2\pi R$ and used the formula,
\begin{equation}
\sum_{n=-\infty}^{\infty}{1\over x^2+(n+a)^2}={\pi\over x}
{\sinh(2\pi x)\over \cosh(2\pi x)-\cos(2\pi a)}.
\end{equation}
Let us note that summing up all the Kaluza-Klein modes is consistent with
the gauge invariance for the direction of the extra dimension. By integrating it 
with respect to $p_E$, we immediately have
\begin{equation}
\sum_{n=-\infty}^{\infty}{\rm ln}\left[p_E^2 +\left({n+Qa-{\delta\over 2}\over R}\right)^2\right]
={\rm ln}\left[\cosh(L p_E)-\cos\left(2\pi\left(Qa-{\delta\over 2}\right)\right)\right].
\end{equation}
It can be shown that the integration
constant does not depend the order parameter $a$, so that we have set it to be zero.

We second perform the $4$-dimensional momentum integration,
\begin{eqnarray}
F_5(Qa, \delta, \tLambda)&=&{1\over 2L^5}{2\pi^2\over \Gamma(2)(2\pi)^4}\int_0^{\tLambda}
d\tpe~\tpe^3 ~{\rm ln}
\left[\cosh \tpe - \cos\left(2\pi \left(Qa-{\delta\over 2}\right)\right)\right]\label{shiki2}\\
&=&{1\over (4\pi )^2L^5}\Biggl[
-6\left({\rm Li}_5(\e^{2\pi i (Qa-{\delta\over 2})})+c.c.\right)
+6\left({\rm Li}_5(\e^{2\pi i (Qa-{\delta\over 2})-\tLambda})+c.c.\right)\nonumber\\
&+&6\tLambda\left({\rm Li}_4(\e^{2\pi i (Qa-{\delta\over 2})-\tLambda})+c.c.\right)
+3\tLambda^2\left({\rm Li}_3(\e^{2\pi i (Qa-{\delta\over 2})-\tLambda})+c.c.\right)
\nonumber\\
&+&\tLambda^3\left({\rm Li}_2(\e^{2\pi i (Qa-{\delta\over 2})-\tLambda})+c.c.\right)
\Biggr],
\label{shiki2new}
\end{eqnarray}
where the dimensionless integration variable have been defined 
by $\tpe\equiv L p_E$ in Eq. (\ref{shiki2}) and $\tLambda=L\Lambda$, and we have 
used the Polylogarithm
defined in Eq. (\ref{shiki3}) in Appendix. Here we have ignored constant
terms that do not depend on the order parameter $a$.
The derivation of Eq. (\ref{shiki2new}) is given in Appendix.

In Eqs. (\ref{shiki2}) and (\ref{shiki2new}) we have 
introduced a dimensionless parameter $\tLambda$ which relates the
$4$-dimensional cutoff $\Lambda$ and the energy scale of the extra dimension $L^{-1}$ as
\begin{equation}
\tLambda = L\Lambda={\Lambda\over 1/L}\equiv \xi.
\label{interrelation}
\end{equation}
The parameter $\xi$ in Eq. (\ref{interrelation}) plays an important role 
in the low-energy physics. Let us call it as {\it interrelation} between a $4$-dimensional
physics and the extra dimension. Here we notice that
$\tLambda$ stands for not only the cutoff, but also the contribution of the Kaluza-Klein 
mode, depending on the scale of $\Lambda$. Namely, the latter point of view is 
crucial for the interrelation, which will be discussed in the section $3$, so that we
shall use the different notation $\xi$ when we emphasize the interrelation such as the 
calculation of the Higgs mass.

Originally the $5$-dimensional dynamics is out of control due to the nonrenormalizabilty. 
A cutoff must be introduced to define the theory, and it lies in certain energy scale
though it is unknown where it should be. One does 
not know how much we should take account of the quantum 
correction from the Kaluza-Klein mode in order to determine the low-energy physics.
At present, the discovery of the Higgs boson has been 
reported \cite{lhc1, lhc2} and we expect the consistent cutoff with LHC result
must lie in certain energy scale. Then the interrelation tells us
how much quantum correction from the Kaluza-Klein mode one should take into account 
in order for the cutoff to be consistent with the LHC result. At the one-loop level, the
effective potential is written in terms of the interrelation and, as we 
will see concretely later, the interrelation becomes manifest through the 
Higgs mass.

The first term in the right hand side of Eq. (\ref{shiki2new}) is well-known and has been 
obtained in the past calculation \cite{klimy}. One observes that all
the terms except for the first term have a remarkable combination
of $\xi$ and the order parameter\footnote{The boundary 
condition $\delta$ of the field is not essential in this discussion, so that we have ignored it.},
\begin{equation}
\e^{2\pi i Qa-\xi}.
\label{dep}
\end{equation}
The combination is resulted by respecting the gauge invariance for the direction of
the extra dimension, that is, the periodicity of the order parameter $a$ and introducing 
the $4$-dimensional cutoff in the momentum integration (\ref{shiki2}). 
The potentially dangerous order parameter 
dependent divergence disappears as $\xi(=\tLambda)$ goes to infinity thanks to the
exponential damping. Let us  note that the exponential behavior of the 
cutoff (\ref{dep}) never appears in the dimensional regularization.

The combination (\ref{dep}) is traced back to Eq.  (\ref{shiki7}). By setting $\delta =0$, it is 
rewritten as 
\begin{eqnarray}
\sum_{n=-\infty}^{\infty} {2p_E R^2\over (Rp_E)^2+(n+Qa)^2}
&=&L\times {\sinh(Lp_E)\over \cosh(Lp_E)-\cos(2\pi Qa)}\nonumber\\
&=&L\times 
\left(1+\left\{{\e^{2\pi i Qa -\tilde p_E}\over 1-\e^{2\pi i Qa -\tilde p_E}}+c.c.
\right\}\right).
\label{shiki8}
\end{eqnarray}
Then the relevant quantity is obtained by the integral of the form,
\begin{equation}
I(\tLambda)\equiv \int_0^{\tLambda}dy~f(y)~\e^{i{\bar a} -y},
\end{equation}
where the function $f(y)$ is an $n$-th polynomial, $f(y)=
\displaystyle{\sum_{k=1}^n a_ky^k}$.
The above integral is evaluated as
\begin{equation}
I(\tLambda)=F(0)\e^{i{\bar a}}-F(\tLambda)\e^{i{\bar a}-\tLambda}=
I(\infty)-F(\tLambda)\e^{i{\bar a}-\tLambda}.
\label{newshiki1}
\end{equation}
Here we have defined 
\begin{equation}
F(y)\equiv \sum_{m=0}^n f^{(m)}(y).
\end{equation}
The first term in Eq. (\ref{newshiki1}) corresponds to the well-known
finite term obtained in the past calculation. It is interesting to note
that the ultraviolet limit of the function $I(\tLambda)$ is
evaluated at the infrared point of the integration variable $y=0$
for another function $F(y)$. This is a notable feature in the
scenario of the gauge-Higgs unification.

The effective potential is a special
quantity in the gauge-Higgs unification because of the combination
$\e^{2\pi iQa-\xi}$ at least at the one-loop level, which 
is never observed in the usual quantum field theory. Once we 
recognize this point, one immediately realizes that quantity other than 
this type does not possess such the combination and hence the 
finiteness. As we will see below, the Higgs mass also has the same 
combination.

Now let us proceed to the Higgs mass, which is obtained by the second derivative 
of the effective potential at the vacuum denoted by $a=a_0$,
\begin{equation}
m_H^2\equiv {\del^2V_{eff}\over \del\vev{A_y^{6}}^2}\Bigg|_{vac}
=(gR)^2{\del^2V_{eff}\over \del a^2}\Bigg|_{a=a_0}.
\label{shiki10}
\end{equation}
The structure of the second derivative of the effective potential can be seen 
from Eq. (\ref{shiki2new}) by
\begin{eqnarray}
{\del^2V_{eff}\over \del a^2}&\propto &
{\del^2 F(Qa,\delta, \xi)\over \del a^2}
\nonumber\\
&\propto&-6\left({\rm Li}_3(\e^{2\pi i (Qa-{\delta\over 2})})+c.c.\right)
+6\left({\rm Li}_3(\e^{2\pi i (Qa-{\delta\over 2})-\xi})+c.c.\right)\nonumber\\
&+&6\xi\left({\rm Li}_2(\e^{2\pi i (Qa-{\delta\over 2})-\xi})+c.c.\right)
+3\xi^2\left({\rm Li}_1(\e^{2\pi i (Qa-{\delta\over 2})-\xi})+c.c.\right)
\nonumber\\
&+&\xi^3\left({\rm Li}_0(\e^{2\pi i (Qa-{\delta\over 2})-\xi})+c.c.\right).
\end{eqnarray}
As stated before, we confirm that the Higgs mass also possesses the same 
combination, $\e^{2\pi i Qa-\xi}$ as that of the effective potential.

If  one takes the infinite limit of $\xi$, only the
first finite term in Eq. (\ref{shiki2new}) is survived to reproduce the well-known 
expression for the Higgs mass. In order to make discussions on the 
interrelation concretely, we need to consider 
models explicitly, which will be given in the next section.
\section{Higgs as Interrelation between 4 and extra  dimensions}
Let us introduce a set of matter.
We follow the studies of the gauge-Higgs unification made in 
the past \cite{finitet, marutake, hty}, in which
we have introduced the fermions and bosons satisfying the periodic boundary condition
$(\eta=1)$ and antiperiodic boundary 
condition $(\eta=-1)$, and whose representations under the $SU(3)$ gauge
group are the adjoint and fundamental ones. We denote their flavor numbers by
\begin{equation}
(N_F^{adj(+)}, N_F^{fd(+)}, N_S^{adj(+)}, N_S^{fd(+)}),~~
(N_F^{adj(-)},N_F^{fd(-)}, N_S^{adj(-)}, N_S^{fd(-)}).
\end{equation} 
Here the $N_{F(S)}^{adj(fd)}$ stands for the number of the fermion (scalar) belonging to 
the adjoint (fundamental) representation under the $SU(3)$ gauge group.
The $\pm$ sign associated with $N_{F(S)}^{adj(fd)}$ is the periodicity 
of the matter field, $\eta=\pm 1$.

Recalling the equation (\ref{shiki9}), the effective 
potential with these types of matter fields are given by
\begin{eqnarray}
&&V_{eff}^{total}={1\over (4\pi)^2L^5}\Bigl[ 
(-1)^0 3J^{adj(+)}+(-1)^1 4N_F^{adj(+)}J^{adj(+)}
+(-1)^1 4N_F^{fd(+)}J^{fd(+)}\nonumber\\
&&+(-1)^0 2N_S^{adj(+)}J^{adj(+)}
+(-1)^0 2N_S^{fd(+)}J^{fd(+)}
+(-1)^1 4N_F^{adj(-)}J^{adj(-)}\nonumber\\
&&+(-1)^1 4N_F^{fd(-)}J^{fd(-)}
+(-1)^0 2N_S^{adj(-)}J^{adj(-)}
+(-1)^0 2N_S^{fd(-)}J^{fd(-)}\Bigr],
\end{eqnarray}
where the first term is the contribution from the gauge bosons, and we have defined
\begin{eqnarray}
J^{adj(+)}&\equiv &F^{\infty}(2a, 0)+F^{\xi}(2a, 0,\xi)
+2(F^{\infty}(a, 0)+F^{\xi}(a, 0,\xi)),\\
J^{adj(-)}&\equiv& F^{\infty}(2a, 1)+F^{\xi}(2a, 1,\xi)
+2(F^{\infty}(a, 1)+F^{\xi}(a, 1,\xi)),\\
J^{fd(+)}&\equiv & F^{\infty}(a, 0)+F^{\xi}(a, 0, \xi),\\
J^{fd(-)}&\equiv &F^{\infty}(a, 1)+F^{\xi}(a, 1,\xi)
\end{eqnarray}
and 
\begin{eqnarray}
F^{\infty}(x, \delta)&=&-6\left({\rm Li}_5(\e^{2\pi i ({x\over 2}-{\delta\over 2})})+c.c.\right),
\label{tsuika1}\\
F^{\xi}(x, \delta,\xi)
&=&
6\left({\rm Li}_5(\e^{2\pi i ({x\over 2}-{\delta\over 2})-\xi})+c.c.\right)
+6\xi\left({\rm Li}_4(\e^{2\pi i ({x\over 2}-{\delta\over 2})-\xi})+c.c.\right)
\nonumber\\
&+&
3\xi^2\left({\rm Li}_3(\e^{2\pi i ({x\over 2}-{\delta\over 2})-\xi})+c.c.\right)+
\xi^3\left({\rm Li}_2(\e^{2\pi i ({x\over 2}-{\delta\over 2})-\xi})+c.c.\right).
\label{tsuika2}
\end{eqnarray}
The shape of the effective potential is determined once we fix the number of flavor 
and $\xi$. In the limit 
of $\xi\rightarrow \infty$,  the $F^{\xi}(x, \delta, \xi)$
vanishes, and the effective potential is given by the function $F^{\infty}(x, \delta)$ 
alone, which is consistent with the results obtained in the past calculation. The effective potential
vanishes at $\xi=0$, as seen from Eqs. (\ref{tsuika1}) and (\ref{tsuika2}).

Let us also give the second derivative of the effective potential which
is necessary for the calculation of the Higgs mass by Eq. (\ref{shiki10}).
\begin{eqnarray}
{\del^2V_{eff}^{total}\over \del a^2}
&=&{(2\pi)^2\over (4\pi)^2L^5}(-1)
\Biggl[
(-1)^03J^{adj(+)}_H+(-1)^14N_F^{adj(+)}J_H^{adj(+)}
+(-1)^14N_F^{fd(+)}J_H^{fd(+)}\nonumber\\
&+&(-1)^02N_S^{adj(+)}J_H^{adj(+)}
+(-1)^02N_S^{fd(+)}J_H^{fd(+)}
+(-1)^14N_F^{adj(-)}J_H^{adj(-)}\nonumber\\
&+&(-1)^14N_F^{fd(-)}J_H^{fd(-)}
+(-1)^02N_S^{adj(-)}J_H^{adj(-)}
+(-1)^02N_S^{fd(-)}J_H^{fd(-)},
\Biggr],
\label{shiki13}
\end{eqnarray}
where we have defined 
\begin{eqnarray}
J^{adj(+)}_H&\equiv &F^{\infty}_H(2a, 0)+F^{\xi}_H(2a, 0,\xi)
+{1\over 4}\times 2(F^{\infty}_H(a, 0)+F^{\xi}_H(a, 0,\xi)),\\
J^{adj(-)}_H&\equiv& F^{\infty}_H(2a, 1)+F^{\xi}_H(2a, 1,\xi)
+{1\over 4}\times 2(F^{\infty}_H(a, 1)+F^{\xi}_H(a, 1,\xi)),\\
J^{fd(+)}_H&\equiv & {1\over 4}\left(F^{\infty}_H(a, 0)+F^{\xi}_H(a, 0, \xi)\right),\\
J^{fd(-)}_H&\equiv &{1\over 4}\left(F^{\infty}_H(a, 1)+F^{\xi}_H(a, 1,\xi)\right)
\end{eqnarray}
and 
\begin{eqnarray}
F^{\infty}_H(x, \delta)&=&-6\left({\rm Li}_3(\e^{2\pi i ({x\over 2}-{\delta\over 2})})+c.c.\right),
\label{shiki11}\\
F^{\xi}_H(x, \delta,\xi)
&=&
6\left({\rm Li}_3(\e^{2\pi i ({x\over 2}-{\delta\over 2})-\xi})+c.c.\right)
+6\xi\left({\rm Li}_2(\e^{2\pi i ({x\over 2}-{\delta\over 2})-\xi})+c.c.\right)
\nonumber\\
&+&
3\xi^2\left({\rm Li}_1(\e^{2\pi i ({x\over 2}-{\delta\over 2})-\xi})+c.c.\right)+
\xi^3\left({\rm Li}_0(\e^{2\pi i ({x\over 2}-{\delta\over 2})-\xi})+c.c.\right).
\label{shiki12}
\end{eqnarray}
The $F^{\xi}_H(x, \delta, \xi)$ vanishes for $\xi\rightarrow \infty$
to reproduce the old results for the Higgs mass, which 
is given by Eq. (\ref{shiki11}). At $\xi=0$, the Higgs 
mass vanishes, as seen from Eqs. (\ref{shiki11}) and (\ref{shiki12}). 
The Higgs mass is given by \footnote{The models of the gauge-Higgs
unification in this paper do not predict the correct the Weinberg angle, and we 
implicitly assume that we have made the prescription done, for 
example, in \cite{hhky}, so that the $4$-dimensional gauge coupling becomes free parameter 
and that its size is of order of one. }
\begin{equation}
m_H^2=(gR)^2{\del^2V_{eff}\over \del a^2}\Bigg|_{a=a_0}
={(2\pi gR)^2\over (4\pi)^2L^5}H(Qa_0, \delta, \xi)
={g_4^4\over(8\pi^2)^2}\left({v\over a_0}\right)^2H(Qa_0, \delta,\xi),
\end{equation}
where we have used the relation $v=a_0/(g_4R)$ followed from the weak
gauge boson mass $M_W=a_0/(2R)$, and we have defined $H(Qa, \delta,\xi)$ by 
the expression aside from the factor $(2\pi)^2/((4\pi)^2L^5)$ in Eq. (\ref{shiki13}). The
$4$-dimensional gauge coupling is defined by $g_4\equiv g/\sqrt{L}$. The 
value of the Higgs mass is determined by putting the values of $a_0, \xi$ and 
the number of flavor.

Now let us first study the case,  where the matter content is given by
\begin{equation}
\mbox{model~A}:\left\{
\begin{array}{cc}
(N_F^{adj(+)}, N_F^{fd(+)}, N_S^{adj(+)}, N_S^{fd(+)})=&(2, 2, 0, 0),\\[0.2cm]
(N_F^{adj(-)},N_F^{fd(-)}, N_S^{adj(-)}, N_S^{fd(-)})=&(2, 2, 0, 3).
\end{array}\right. 
\end{equation}
We first present the typical shape of the effective potential 
for $\tLambda\rightarrow \infty$ in Fig. $1$. 
\begin{center}
\begin{figure}[ht]
\begin{center}
\includegraphics[width=10cm]{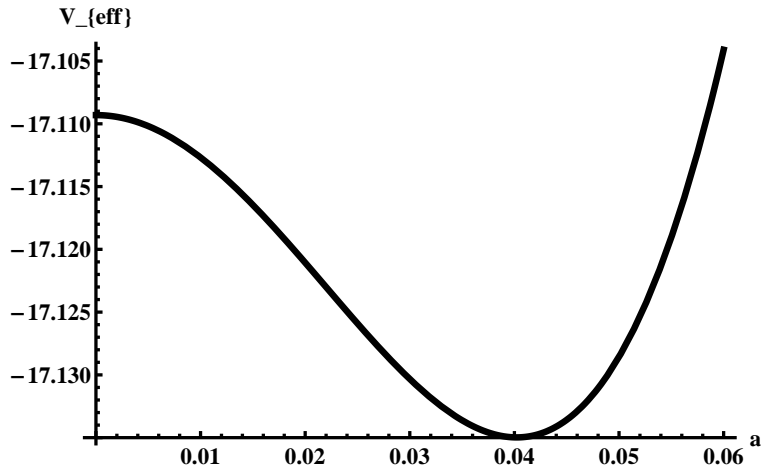}
\caption{The  shape of the effective potential in the 
limit of $\tLambda\rightarrow\infty$ for the model A. 
The global minimum is located at $a_0=0.0402199$. }
\end{center}
\end{figure}
\end{center}
The global minimum is
located at $a_0=0.0402199$ and the $SU(2)\times U(1)$
gauge symmetry breaks down to $U(1)_{em}$. By using the vacuum expectation 
value $a_0$, the Higgs mass in the same limit is 
calculated as $m_H/g_4^2=130.222$[GeV]. It has been known that
the matter content is crucial for obtaining the sufficiently heavy 
Higgs mass \cite{hty}.

Now we turn on the cutoff $\tLambda(=\xi)$. The shape of the effective potential is changed 
according to the value of $\tLambda$, so that the position of the global minimum 
is also changed. We show the behavior of $a_0$ with
respect to $\tLambda$ in Fig. $2$. The gauge symmetry is correctly broken, that is, $a_0\neq 0, 1$ for 
the range of $\tLambda$ we have studied \footnote{At $\tLambda=0$, the effective 
potential vanishes, so that the position of the global minimum in the limit is unclear.}. 
The magnitude of $\tLambda$ for $\tLambda~\rough{>}~10$ almost saturates the values 
obtained in the limit of $\tLambda\rightarrow \infty$.  
\begin{center}
\begin{figure}[ht]
\begin{center}
\includegraphics[width=11cm]{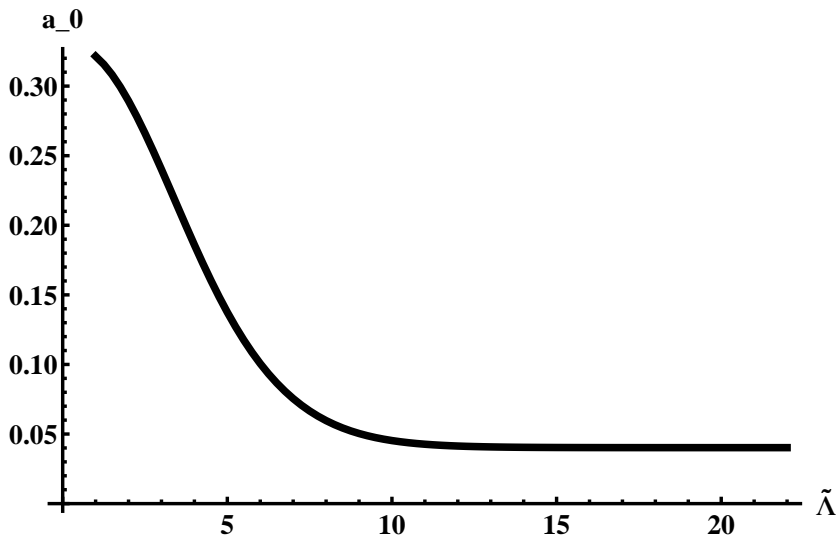}
\caption{The behavior of the order parameter $a$ with 
respect to $\tLambda$ for the model A.}
\end{center}
\end{figure}
\end{center}

Let us next depict  the behavior of the Higgs mass with respect to the
interrelation $\xi={\Lambda\over 1/L}$ in Fig. $3$. 
\begin{center}
\begin{figure}[ht]
\begin{center}
\includegraphics[width=11cm]{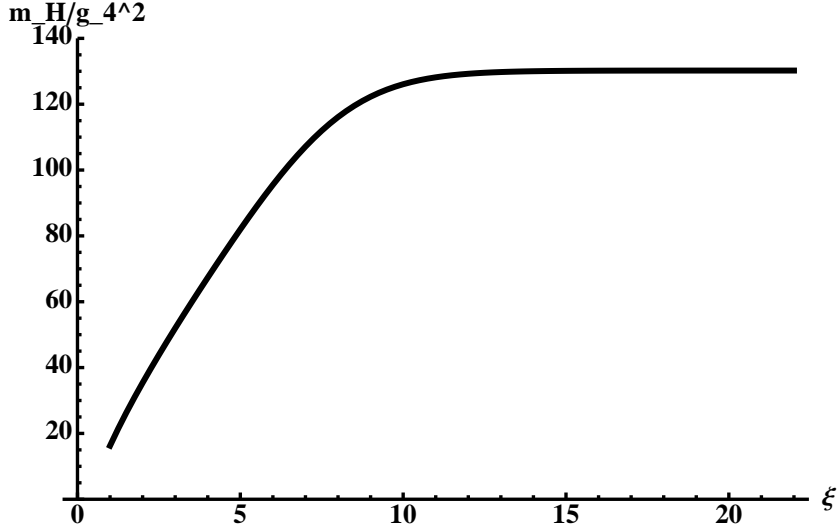}
\caption{The behavior of the Higgs mass with respect to $\xi={\Lambda\over1/L}$ for 
the model A. The asymptotic value of the Higgs 
mass is about $130$[GeV].}
\end{center}
\end{figure}
\end{center}
We observe that the Higgs mass becomes larger as $\xi$ is 
larger and for $\xi~\rough{>}~10$ the Higgs mass is almost saturates the value 
obtained in the limit of $\xi\rightarrow \infty$. On the other 
hand, for $1~\rough{<}~\xi ~\rough{<}~8$, the Higgs mass grows almost linearly with 
respect to $\xi$. If we take account of the recently reported Higgs mass, $126$ [GeV]
at LHC \cite{lhc1, lhc2}, we obtain a bound on $\xi$. It is given 
by $\xi={\Lambda\over1/L}~\rough{>}~10$, which implies that the $4$-dimensional cutoff $\Lambda$
must satisfy $\Lambda~\rough{>}~10L^{-1}$. The value of the Higgs mass is
smoothly connected to zero for $\xi\rightarrow 0$ as far as our numerical
analyses are concerned.

We can also understand the behavior of the Higgs mass with 
respect to $\xi$ by the first derivative of $F^{\xi}_H (x, \delta, \xi)$, which
controls the Higgs mass essentially. It is given by
\begin{eqnarray}
{\del F_H^{\xi}(x, \delta, \xi)\over \del \xi}
&=&-\xi^3 \sum_{n=1}^{\infty} n 
\left(\e^{2\pi i n({x\over 2}-{\delta\over 2})-n\xi}+c.c.\right)\nonumber\\
&=&-\xi^3\Biggl[
{\e^{2\pi i ({x\over 2}-{\delta\over 2})-\xi}
\over \left(1-\e^{2\pi i({x\over 2}-{\delta\over 2})-\xi}\right)^2}
+c.c.\Biggr].
\end{eqnarray}
For large value of $\xi$, due to the exponential damping factor, the 
first derivative vanishes, so that the value of the
Higgs mass becomes constant. This corresponds to the flat behavior 
in Fig. $3$. When $\xi$ becomes larger from zero, the $\xi^3$ starts to 
control the behavior of the Higgs mass. This gives the almost 
linear growth of the Higgs mass with respect to $\xi$ in Fig. $3$.

Let us discuss the interrelation $\xi= {\Lambda\over 1/L}$ which 
is manifest through the Higgs mass. If the $4$-dimensional 
cutoff $\Lambda$ is smaller than the the scale of the extra
dimension, $\Lambda < L^{-1}$, the Kaluza-Klein modes can not 
be excited in the $4$ dimensions. Since the Higgs mass
is essentially generated by the quantum 
effect of the Kaluza-Klein mode, the Higgs mass is tiny enough 
for the region of the scale $\xi~\rough{<}~1$. As the 
cutoff $\Lambda$ becomes larger,  the Kaluza-Klein modes 
can start to excite and contribute to the Higgs mass, so that it 
gradually becomes heavier. This corresponds to the slop in the region 
of $1~\rough{<}~\xi~\rough{<}~8$. When the $\Lambda$ becomes
large further, $L^{-1}\leq \Lambda$, the Kaluza-Klein 
modes can be excited fully enough to yield the Higgs mass corresponding to 
the flat part. The behavior of the Higgs mass clearly shows the interrelation 
between the effect of the $4$-dimensional cutoff and the 
physics in $5$ dimensions, that is, the Kaluza-Klein mode.

As an illustration, let us also consider the two more cases, where the matter 
contents are given by
\begin{eqnarray}
&&\mbox{model~B}:\left\{
\begin{array}{cc}
(N_F^{adj(+)}, N_F^{fd(+)}, N_S^{adj(+)}, N_S^{fd(+)})=&(3, 2, 0, 0),\\
(N_F^{adj(-)},N_F^{fd(-)}, N_S^{adj(-)}, N_S^{fd(-)})=&(4, 1, 1, 3).
\end{array}\right.
\\
&&\mbox{model~C}:\left\{
\begin{array}{cc}
(N_F^{adj(+)}, N_F^{fd(+)}, N_S^{adj(+)}, N_S^{fd(+)})=&(3, 4, 0, 0),\\
(N_F^{adj(-)},N_F^{fd(-)}, N_S^{adj(-)}, N_S^{fd(-)})=&(5, 1, 2, 4).
\end{array}\right.
\end{eqnarray}
In the limit of $\tLambda\rightarrow \infty$, the Higgs mass in the
model B (C) is $186.694 (168.096)$[GeV], where the
order parameter at the vacuum is given by $a_0=0.0285365 (0.0436442)$.

We turn on the cutoff $\tLambda$ and depict 
the behavior of the order parameter $a_0$ in Fig. $4$ for the models B and C.
For the range of $\tLambda$, we have studied the gauge symmetry is broken correctly.
In Fig. $5$, we show the behaviors of the Higgs mass for the two models.
For the model B (C), if we take account of the LHC result of the Higgs 
mass $126$ [GeV], we obtain  $\xi={\Lambda\over1/L}~\rough{>}~5.7 (6.26)$, which 
implies $\Lambda~\rough{>}~5.7 (6.26)L^{-1}$.
\begin{center}
\begin{figure}[ht]
\begin{center}
\includegraphics[width=12cm]{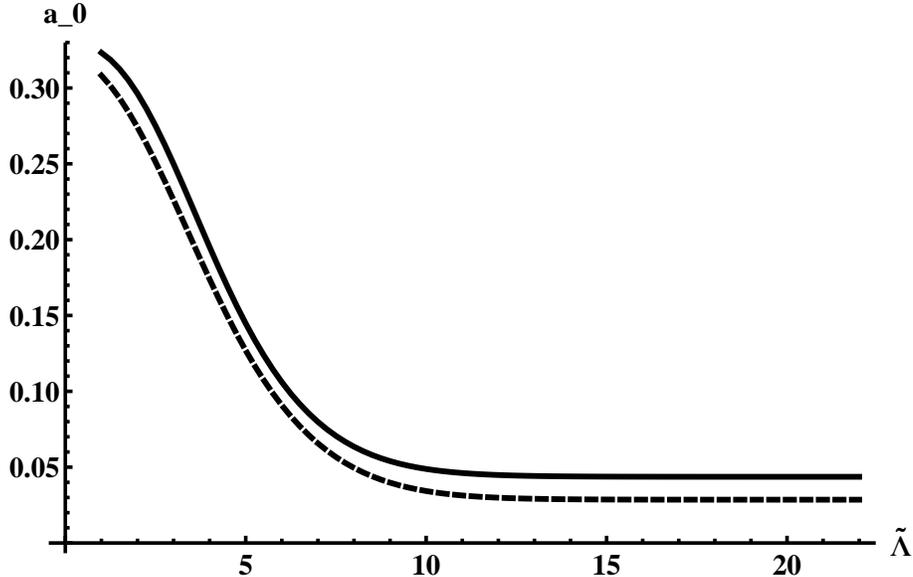}
\caption{The behavior of the order parameter $a$ with respect to $\tLambda$. 
The dotted (solid) line stands for the case of model C (B). }
\end{center}
\end{figure}
\end{center}
\begin{center}
\begin{figure}[ht]
\begin{center}
\includegraphics[width=12cm]{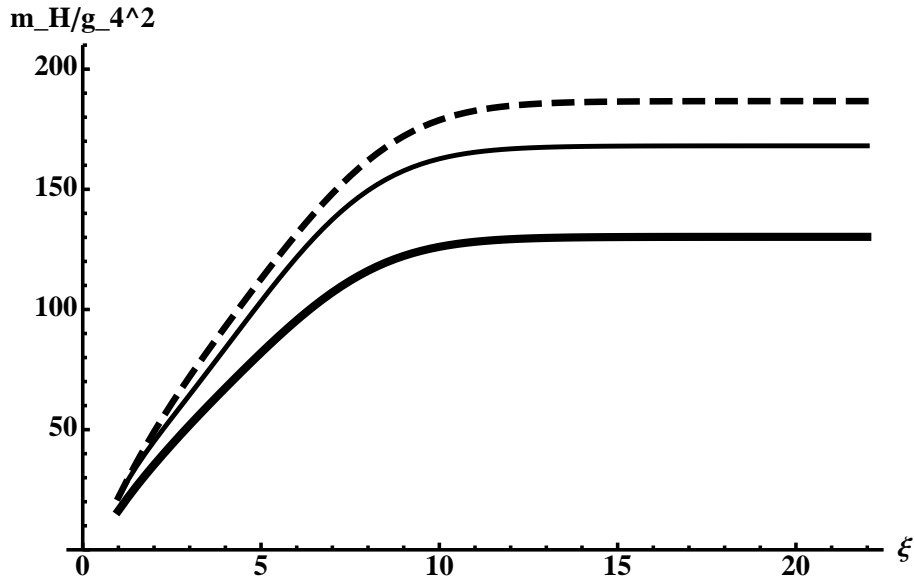}
\caption{The behavior of the Higgs mass with respect 
to $\xi={\Lambda\over 1/L}$. The thick (thin, dashed) line is the 
case for the model A (C, B). The asymptotic value of the Higgs mass for each model is
about $130 (168, 186)$[GeV] for model A (C, B).}
\end{center}
\end{figure}
\end{center}
\section{Conclusions}
We have evaluated the one-loop effective potential and the Higgs mass in the scenario of the 
gauge-Higgs unification by introducing the $4$-dimensional cutoff $\Lambda$ in order to
control the ultraviolet effect. It is clarified how much Kaluza-Klein mode
appeared in $4$ dimensions contributes to the effective potential and the
Higgs mass thanks to the cutoff. The effective potential and the Higgs mass 
depend on both the order parameter $a$ and $\xi\equiv {\Lambda\over 1/L}$ 
through the remarkable combination $\e^{2\pi i Qa-\xi}$. Due to
the exponential damping, the well-known terms that obtained in
the past calculations are reproduced in the limit of $\xi\rightarrow\infty$.

The parameter $\xi={\Lambda\over 1/L}$ stands for the
interrelation, which is, in particular, concretized through 
the Higgs mass. We have presented the three models in order to study the interrelation. 
We have obtained the behaviors of the Higgs mass with respect 
to $\xi={\Lambda\over 1/L}$. The 
behavior shows the interrelation between 
the $4$ and the extra dimensions. For the smaller cutoff $\Lambda$, the Kaluza-Klein 
excitations are suppressed in $4$ dimensions, so that the Higgs 
mass, which essentially originates from the quantum effect of the 
Kaluza-Klein mode, is suppressed as well. As the cutoff $\Lambda$ becomes 
larger, the excitations can be allowed to generate      
the Higgs mass gradually and for the certain large value 
of $\Lambda$, the Higgs mass approaches to 
the value obtained in the limit of $\Lambda\rightarrow \infty$, which
means that the quantum correction in the extra dimension is fully incorporated. The 
interrelation is manifest through the Higgs mass, which shows that the $5$-dimensional 
effect dominates for the large $\Lambda$, while the $4$-dimensional cutoff 
becomes effective for the smaller $\Lambda$.

We have also obtained the bound on $\xi$ by taking 
account  of the LHC result. This, in turn, gives the bound on the ratio between
the $4$-dimensional cutoff $\Lambda$ and the scale 
of the extra dimensions $1/L$.

The combination $\e^{2\pi i Qa-\xi}$ is remarkable 
if we think of the usual logarithm and power behaviors with respect 
to the cutoff in the quantum field theory. The combination
shows that the effective potential and Higgs mass are the special 
quantities in the gauge-Higgs unification. The origin of the combination
may be the gauge invariance in the extra dimension. It is 
interesting to ask whether such the combination still holds beyond the one-loop 
calculation \cite{twoloop} and to investigate the role of the combination 
further. It may shed new light on the gauge-Higgs unification from a point of view
of quantum field theory. Of course, it is important to study nonperturbatively 
the $5$-dimensional gauge theory in a view of the 
interrelation. These will be reported in elsewhere \cite{sotake}.
\begin{center}
{\bf Acknowledgement}
\end{center}
This work is supported in part by a Grant-in-Aid for Scientific Research
(No. 20540274, 25400260 (H.S.), No. 24540291 (K.T.)) from the Japanese 
Ministry of Education, Science, Sports and Culture. 
\vspace*{1cm}
\begin{center}
{\bf \Large Appendix}  
\end{center}
\begin{flushleft}
{\bf Derivation of Eq. (\ref{shiki2new})}
\end{flushleft}
The momentum integration in Eq. (\ref{shiki2}) can be performed analytically. It is 
easy to show that the indefinite integration is carried out as 
\begin{eqnarray}
&&\int dy~y^{3}~{\rm ln}\left(\cosh y -\cos{\bar a}\right)
=\int dy \left({y^4\over 4}\right)'   {\rm ln}\left(\cosh y -\cos{\bar a}\right)\nonumber\\
&=&{y^4\over 4}{\rm ln}\left(\cosh y -\cos{\bar a}\right)-\int dy~
{y^4\over 4}\left(1+\left\{{\e^{i{\bar a}-y}\over {1-\e^{i{\bar a}-y}}}+c.c.\right\}\right)
\nonumber\\
&=&
{y^{4}\over 4}{\rm ln}\left[\cosh y- \cos{\bar a}\right]
-{y^5\over 20}
-{y^{4}\over 4}\left({\rm ln}(1-\e^{i{\bar a}-y})+c.c.\right)\nonumber\\
&&+\int dy~y^3\left({\rm ln}(1-\e^{i{\bar a}-y})+c.c.\right).
\label{shiki14}
\end{eqnarray}
It is straightforward to show that the first three terms in Eq. (\ref{shiki14}) become
\begin{eqnarray}
&&{y^4\over 4}{\rm ln}\left[\cosh y- \cos{\bar a}\right]
-{y^5\over 20}
-{y^4\over 4}\left({\rm ln}[1-\e^{i{\bar a}-y}]+c.c.\right)\nonumber\\
&&={-y^5\over 20}+{y^4\over 4}{\rm ln}\left({\cosh y-\cos{\bar a}\over (1-\e^{-i{\bar a}-y})
(1-\e^{{i\bar a}-y})}\right)\nonumber\\
&&={-y^5\over 20}+{y^4\over 4}{\rm ln}\left({\e^y\over 2}\right)
={y^5\over 5}-{{\rm ln}2\over 4}y^{4}.
\end{eqnarray}
In the second line of Eq. (\ref{shiki14}) we first expand the logarithm 
by \footnote{Note that the mode $n$ in Eqs. (\ref{shiki4}) and (\ref{shiki3}) is different 
from the original Kaluza-Klein mode $n$. We point out that the mode summation (\ref{shiki4}) 
is the same as the one obtained by the Poisson's resummation formula.}
\begin{equation}
{\rm ln}(1-x)=-\sum_{n=1}^{\infty}{x^n\over n},
\label{shiki4}
\end{equation}
and after the partial integration we make use of the Polylogarithm, 
\begin{equation}
{\rm Li}_{s}(z)\equiv \sum_{n=1}^{\infty}{z^n\over n^s}.
\label{shiki3}
\end{equation}
We finally obtain that
\begin{eqnarray}
\int dy~y^{3}~{\rm ln}\left[\cosh y -\cos{\bar a}\right]
&=&{y^5\over 5}-{{\rm ln}~2\over 4}y^4\nonumber\\
&+&y^3\left({\rm Li}_2(\e^{i{\bar a}-y})+c.c.\right)
+3y^{2}\left({\rm Li}_3(\e^{i{\bar a}-y})+c.c.\right)\nonumber\\
&+&6 y\left({\rm Li}_4(\e^{i{\bar a}-y})+c.c.\right)
+6 \left({\rm Li}_5(\e^{i{\bar a}-y})+c.c.\right).
\label{shiki5}
\end{eqnarray}
The first and second terms, which are independent on $\bar a$ and 
something like the cosmological constant. Equipped 
with Eq. (\ref{shiki5}), the momentum 
integration (\ref{shiki2}) is evaluated as Eq. (\ref{shiki2new}).

The momentum integration for 
the case of $M^{D-1}\times S^1/Z_2$ is also carried out in the same manner. It 
is given by
\begin{eqnarray}
&&\int dy~y^{D-2}~{\rm ln}\left(\cosh y -\cos{\bar a}\right)=\int dy~\left({y^{D-1}\over {D-1}}\right)'
~{\rm ln}\left(\cosh y -\cos{\bar a}\right)\nonumber\\
&=&{y^{D-1}\over D-1}{\rm ln}\left(\cosh y-\cos{\bar a}\right)-
\int dy~{y^{D-1}\over D-1}\left(1+\left\{ {\e^{i{\bar a}-y}\over{1-\e^{{i\bar a}-y}}}+c.c.\right\}\right)
\nonumber\\
&=&{y^D\over D}-{{\rm ln}2\over D-1}y^{D-1}\nonumber\\
&+&y^{D-2}\left({\rm Li}_2(\e^{i{\bar a}-y})+c.c.\right)
+(D-2)y^{D-3}\left({\rm Li}_3(\e^{i{\bar a}-y})+c.c.\right)\nonumber\\
&+&(D-2)(D-3)y^{D-4}\left({\rm Li}_4(\e^{i{\bar a}-y})+c.c.\right)\nonumber\\
&+&(D-2)(D-3)(D-4)y^{D-5}\left({\rm Li}_5(\e^{i{\bar a}-y})+c.c.\right)+\cdots\nonumber\\
&+&(D-2)(D-3)\cdots(D-(D-2))(D-(D-1))\left({\rm Li}_D(\e^{i{\bar a}-y})+c.c.\right).
\end{eqnarray}
$D=5$ is our case (\ref{shiki5}).
%
%
%

\end{document}